\documentclass{aa}
\usepackage{graphicx}
\usepackage{times}
\usepackage{astron}
\usepackage{psfig}
\begin{document}

\title{Deep BVR photometry of the Chandra Deep Field South from the COMBO-17 survey}

\author{C. Wolf\inst{1} \and S. Dye\inst{1,2} \and M. Kleinheinrich\inst{1} 
\and H.-W. Rix\inst{1} \and K. Meisenheimer\inst{1} \and L. Wisotzki\inst{3} }

\institute{ Max-Planck-Institut f\"ur Astronomie, K\"onigstuhl 17,
            D-69117 Heidelberg, Germany 
       \and Astrophysics Group, Blackett Lab,
            Imperial College, Prince Consort Road, London, U.K.  
       \and Institut f\"ur Physik, Universit\"at Potsdam, Am Neuen
            Palais 10, D-14469 Potsdam, Germany }

\date{Received / Accepted }

\abstract{
We report on deep multi-color imaging ($R_{5\sigma} = 26$) of the Chandra Deep Field South, obtained with the Wide Field Imager (WFI) at the MPG/ESO 2.2\,m telescope on La Silla as part of the multi-color survey COMBO-17. As a result we present a catalogue of 63501 objects in a field measuring $31\farcm5 \times 30\arcmin$ with astrometry and BVR photometry. A sample of 37 variable objects is selected from two-epoch photometry. We try to give interpretations based on color and variation amplitude. 
}
\titlerunning{Deep BVR photometry of the Chandra Deep Field South}
\authorrunning{Wolf et al.}
\maketitle

\section{Introduction}

Deep fields have become a favourite tool of observational cosmology,
particularly in conjunction with the construction of multiwavelength 
datasets. Besides the ubiquitous Hubble Deep Fields, another 
illustrative example is the ROSAT Deep Survey in the 
Lockman Hole (Hasinger et al.\ 1998) and its optical follow-up by
imaging and spectroscopy (e.g., Schmidt et al.\ 1998).
However, full spectroscopic coverage is usually impossible
to obtain, and the reconstruction of distances often has to rely
on photometric redshift estimation. Despite major advances in this
area, there is always a certain degree of degeneracy between object 
classes, such as different galaxy types, and their redshifts. Breaking
this degeneracy is possible only by incorporating independent 
spectroscopic information, such as adding infrared colors
or moving towards finer spectrophotometric resolution. 

Wolf, Meisenheimer, \& R\"oser (2001) and Wolf et al.\ (2001)
demonstrated that the use of medium-band filters 
leads to a substantial gain in classification accuracy and 
discriminative power, as compared to simple broad-band photometry.
In 1999 we initiated a new multicolor survey using the Wide Field
Imager (WFI, Baade et al. 1998, 1999) at the 
MPG/ESO 2.2\,m telescope on La Silla, Chile.
The survey was designed to make full use of the capabilities offered 
by the WFI and at the same time exploit the experience collected in
the course of earlier multicolor projects. 
By incorporating 17 different optical filters into our new survey
the COMBO-17 (\textit{Classifying Objects by Medium-Band
Observations in 17 filters}) project goes a major step beyond the 
traditional multicolor approach. 
It permits confident spectral classification of objects with $R\la 24$ 
into stars, galaxies, quasars, and the recognition of exotic objects,
and it facilitates accurate subclassification, redshifts estimation,
and SED reconstruction for galaxies and quasars. The survey area comprised 
four independent WFI fields amounting to a total area of 1~deg$^2$. 
Overall objectives of COMBO-17 are:
\begin{itemize}
\item The galaxy catalogue will be complete to $R < 24$ and 
      contain $\sim 50\,000$ galaxies with accurate redshifts and 
      spectral classification up to $z \simeq 1.5$, to be used
      to study the evolution of the galaxy luminosity function and
      clustering properties.
\item The quasar sample will contain $\sim 500$ QSOs, with nearly
      uniform completeness over the redshift range $0.5 \la z \la 5$,
      well suited to trace the turnover in QSO evolution.
\item Deep high-resolution $R$-band images with $0\farcs75$ FWHM permit
      morphological classification and gravitational lensing studies.
\end{itemize}

In this paper we focus on the ``Chandra Deep Field South'' \cite{Gia00} 
subject to a deep X-ray exposure by the Chandra
satellite observatory and target of a wide range of
multiwavelength studies from several groups. It is particularly worth
noting that the COMBO-17 field of view contains the Chandra field entirely
facilitating the efficient combination of optical and X-ray data.
In this paper, we present a first catalogue of optically selected sources 
down to $R$-band magnitudes of 
$R \la 26$. This catalogue is publically available over the world-wide web.

\section{Observations}

We have imaged the Chandra Deep Field South in all 17 filters for
the COMBO-17 project using the Wide Field Imager (WFI, Baade et al. 1998,
1999) at the MPG/ESO 2.2\,m telescope on La Silla, Chile. The WFI is a
mosaic camera consisting of eight 2k $\times$ 4k CCDs with $\sim 67$ million
pixels in total, a pixel scale of $0\farcs238$ and a field of view of
$33\arcmin \times 34\arcmin$. The CCDs are rather blue sensitive and
some of them are cosmetically suboptimal since they are only of engineering grade.

Here we discuss data obtained in the broad-band WFI filters $B$, $V$ 
and $R$ (see Table\,\ref{obslog} for a brief observing log). 
Our data encompass a total exposure time of 5000 sec in $B$ and
8400 sec in $V$ with seeing on the order of $\sim 1\farcs2$ and
altogether 23700 sec in $R$ with $\sim 0\farcs75$ mean PSF. Besides long
exposures for efficient light gathering, we included short exposures
for the photometry of brighter objects, in particular to avoid
saturation of our brighter standard stars.

The long exposures followed a dither pattern with ten
telescope pointings spread by $\Delta \alpha$, $\Delta \delta < \pm
72\arcsec$. This dither pattern is motivated by the intent to close
the gaps in the CCD mosaic, but limited by the requirement of keeping 
field rotation at a minimum. Due to the gaps in the CCD mosaic the 
effective exposure time varies within the field. However, dithering
was performed such, that each position on the sky falls onto a CCD
in at least eight out of ten exposures, while 97\% of the area is
always recorded in every image. 

Twilight flatfields were obtained with offsets of $10''$ between
consecutive exposures. Exposure times ranged between 0.5 and 100 seconds
per frame (note that the WFI shutter design allows exposures as short 
as 0.1 seconds without causing significant spatial variations in the 
illumination across the CCD mosaic \cite{Wackermann}).

\begin{table}
\caption{Observing log for $B$, $V$ and $R$ exposures on the Chandra Deep
Field South as part of the COMBO-17 survey. \label{obslog} }
\begin{tabular}{llll}
Epoch (UT) & Band & Seeing & Exposure (sec) \\ 
\noalign{\smallskip} \hline \noalign{\smallskip} 
1999, Oct 10 & V & $1\farcs0 \ldots 1\farcs4$ & $14\times 600$ + $3\times 20$ \\ 
1999, Oct 13 & B & $1\farcs0 \ldots 1\farcs4$ & $10\times 500$ + $3\times 20$ \\ 
1999, Oct 19 & R & $0\farcs6 \ldots 0\farcs9$ & $25\times 420$ \\ 
1999, Oct 20 & R & $0\farcs6 \ldots 0\farcs9$ & $11\times 420$ + $1\times 370$ \\ 
\noalign{\smallskip}
2000, Feb 6 & R & $0\farcs6 \ldots 0\farcs9$ & $6\times 500$ + $2\times 60$ \\
2000, Feb 8 & R & $0\farcs6 \ldots 0\farcs9$ & $10\times 500$ + $1\times 200$ \\ 
\noalign{\smallskip} \hline
\end{tabular}
\end{table}

We have established our own set of tertiary standard stars based on 
\emph{spectrophotometric} observations, mainly in order to achieve a 
homogeneous photometric calibration for all 17 WFI filter bands. Two stars 
of spectral types G--F and magnitudes $B_J \simeq 16$ were selected in
each COMBO-17 field, drawn from the Hamburg/ESO Survey database of
digital objective prism spectra \cite{wisotzki:2000}.  
The spectrophotometric observations for the
Chandra Deep Field South were conducted at La Silla on Oct 25, 1999,
using the Danish 1.54\,m telescope equipped with DFOSC. A wide ($5''$)
slit was used for the COMBO-17 standards as well as for the external
calibrator, in this case the HST standard HD~49798 \cite{bohlin:1992}. 
Two exposures of 45~min were taken of each star,
one with the blue-sensitive grism 4 covering the range $\lambda =
3400$--7400\,\AA , and one with the red-sensitive grism 5 covering
$\lambda > 5200$\,\AA.

The spectra were reduced by standard procedures and have a final
signal-to-noise ratio of $> 30$ per pixel except very near to the low-
and high-wavelength cutoffs. The agreement between spectra in the
substantial overlap in wavelength between the two grisms is excellent,
confirming that contamination from second order was negligible. The
absolute spectrophotometric accuracy, estimated from comparing several
spectra of the external calibrator HD~49798 obtained during the entire
observing run, is better than 10\,\%.

\begin{figure}
\centerline{\hbox{
\psfig{figure=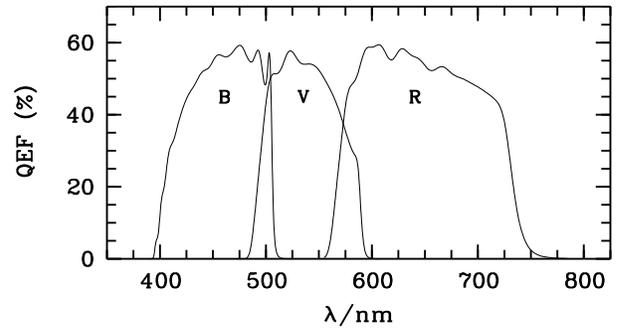,angle=270,clip=t,width=8cm}}}
\caption[ ]{Total system efficiencies for the WFI filters $B$, $V$ and $R$
(including telescope, instrument and detector).  \label{BVRqef}}
\end{figure}

\section{Data reduction}

All procedures used for the data reduction are based on the MIDAS package 
and are routinely used at MPIA. An image processing pipeline has been developed 
specifically for dithered WFI survey images by us. It makes intensive use 
of a programmes developed by K. Meisenheimer, H.-J. R\"oser and H. Hippelein 
for the Calar Alto Deep Imaging Survey (CADIS). The pipeline performes basic 
image reduction and includes
standard operations of bias subtraction, CCD non-linearity correction, 
flatfielding, masking of hot pixels and bad columns, cosmic correction 
and subsequent stacking into a deep coadded frame covering the area that
is common to all frames and covers $31\farcm5 \times 30\arcmin$. Details
on the processing will be given in a forthcoming paper when the data
reduction has been completed for all filters and fields (Meisenheimer et al.,
in preparation).

The coadded frame thus obtained is not optimal for photometry since the
flux errors are not sufficiently described by photon noise only. Instead,
flatfield errors and other systematic effects which are locally changing 
on the CCD can only be incorporated into the error analysis by measuring
the photometry on the individual frames, where the object location varies
due to dithering. Combining these individual measurements allows to derive
flux errors from from the scatter among the frames. 

The deep coadded images we use in fact only for object search and visual 
inspection purposes. Objects have been searched only on the R-band 
sum frame, which provides a uniform, sharp PSF with $0\farcs75$ FWHM and
the best signal-to-noise ratio for all known kinds of objects expected
in the field. We used the SExtractor software \cite{BA96} with the recommended
default setups in the parameter file, except for choosing a minimum of
12 significant pixels required for the detection of an object. 
We first search rather deep 
and then clean the list of found objects from those having more than 
$0\fm333$ error in the SExtractor best-guess magnitude. As a result 
we obtained a catalogue of 63501 objects with positions and morphology.
Starting from the known object positions on the coadded R-band frame, 
we transform the positions of the objects onto every single frame and
measure their fluxes on them. 

\begin{figure}
\centerline{\hbox{
\psfig{figure=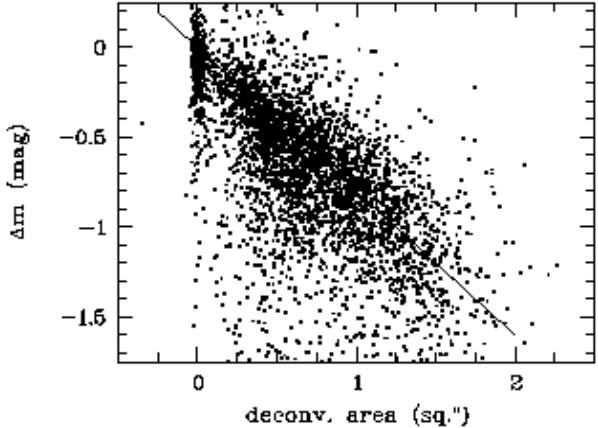,angle=270,clip=t,width=8cm}}}
\caption[ ]{Distribution of aperture correction magnitudes versus
deconvolved area of bright objects. Stars are at zero level while 
extended objects reach down to negative values. The straight line
shows the best fit to the data which has been adopted for a general 
aperture correction that is also applied to faint objects. \label{Aperfit}}
\end{figure}

COMBO-17 is a spectrophotometric 
survey, where color indices are the prime observables entering a process
of classification and redshift estimation later. Therefore, it is necessary 
to choose an optimum way to measure these indices. For ground-based observations 
it is important to avoid that variable observing conditions introduce offsets 
between bands when the observations are taken sequentially. Variable seeing, 
e.g., might influence the flux measurement of star-like and extended objects 
in a different way. 

This requires to assess the seeing point spread function on every frame very 
carefully. Then, we essentially convolve each image to a common \emph{effective} 
point spread function and measure the central surface brightness of each object
in a weighted circular aperture \cite{RM91}. This has the disadvantage that the 
spatial resolution (i.e. the minimum separation of objects neighboring each other) is
limited by the frame with the poorest seeing. Especially, we do not attempt to
separate the fluxes among closely blended objects. For the context of this
paper we use an effective PSF of $1\farcs5$.

The flux calibration is performed by identifying our spectrophotometric standard stars and convolving their spectra with the total system efficiency in the given filter (see Fig.\,\ref{BVRqef}). We then know the physical photon flux we have to assign to them, and establish the flux scale for all objects. Since the spectra of the standard stars have been measured with a 5\arcsec wide slit in good seeing, we are confident that we have collected basically all their light. This implies that the photometry of all stars should be accurate since the standard stars are measured with the same aperture in the images as all other objects. 
\begin{figure*}
\centerline{\hbox{
\psfig{figure=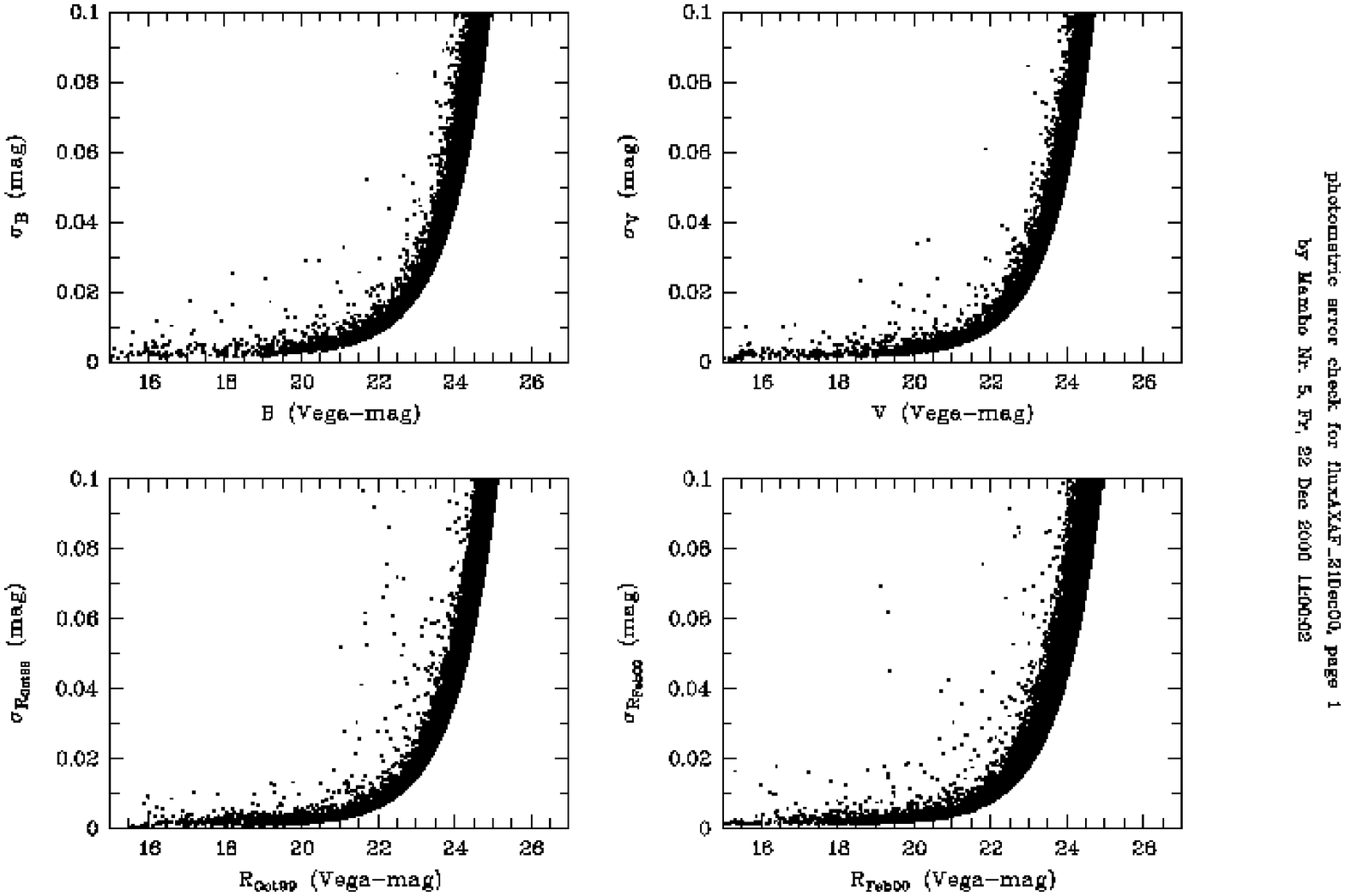,angle=270,clip=t,width=14cm}}}
\caption[ ]{Errors versus magnitudes for all objects with errors below
$0\fm1$ observed in the WFI filters B, V and R (two epochs). \label{errors}}
\end{figure*}

Extended objects however have their fluxes underestimated and therefore we performed a set of photometric runs with apertures increasing in steps to $10\arcsec$ and no weighting functions. At $10\arcsec$ diameter basically all fluxes have already converged. This analysis can not be performed for the faintest objects and therefore we derived an aperture correction function depending on the morphological parameters from the extended objects with $R<22$. From that function an aperture correction is then uniformly estimated for all objects (see Fig.\,\ref{Aperfit}).

The fluxes from individual frames are averaged into a final flux for each object with the error being derived from the scatter. This way, the error does not only take photon noise into account, but further sources of error, such as imperfect flatfielding and uncorrected CCD artifacts. However, we prevent chance coincidences of count rates from pretending unreasonably low errors by using the errors derived from background and photon noise as a lower limit (see Meisenheimer et al., in preparation, for a full discussion of the photometric analysis).

\section{The catalogue and initial results}

\begin{table}
\caption{Format of the object catalogue made available at CDS. \label{CDStab} }
\begin{tabular}{lp{6.5cm}}
Label & Explanation \\
\noalign{\smallskip}
\hline
\noalign{\smallskip}
Seq &       Sequential number\\                          
RA &       Right ascension (J2000), internal accuracy $\pm0\farcs15$ \\                    
DE &       Declination (J2000), internal accuracy $\pm0\farcs15$ \\                        
Bmag &      Aperture magnitude,  scaled to total flux for stars\\                     
e-Bmag &    Mean error (sigma) of Bmag\\    
Vmag &      Aperture magnitude,  scaled to total flux for stars\\                     
e-Vmag &    Mean error (sigma) of Vmag\\    
Rmag &      Aperture magnitude,  scaled to total flux for stars\\                     
e-Rmag &    Mean error (sigma) of Rmag\\ 
ApCmag &    Aperture correction, in units of magnitudes\\
\noalign{\smallskip}
\hline
\end{tabular}
\end{table}

The object catalogue derived from the observations presented contains
positions and \emph{BVR} photometry of 63501 objects selected in $R$ 
within a field of $31\farcm5 \times 30\arcmin$ size. The depth and the seeing
quality of our $R$-band imaging makes this catalogue potentially very
useful for the scientific community. Therefore, the catalogue (format see 
Table\,\ref{CDStab}) is available to the public at Centre de Données astronomiques de 
Strasbourg (CDS, \texttt{http://cdsweb.u-strasbg.fr/}) and on the COMBO-17 survey 
homepage at MPIA (\texttt{http://www.mpia.de/COMBO/}). In the following we discuss 
data quality issues and present a first sample of variable objects identified from 
the two epochs of $R$-band observations.

The quality of our photometry differs for point sources and extended
sources. Essentially, it is a seeing-adaptive central surface brightness 
measurement giving accurate fluxes for point sources while underestimating 
the total flux of extended sources. But since it is performed on the individual 
frames in an optimal seeing-adaptive fashion, it yields more accurate colors
and estimates the errors more realisticly than measurements on a single coadded frame. 

\begin{figure}
\centerline{\hbox{
\psfig{figure=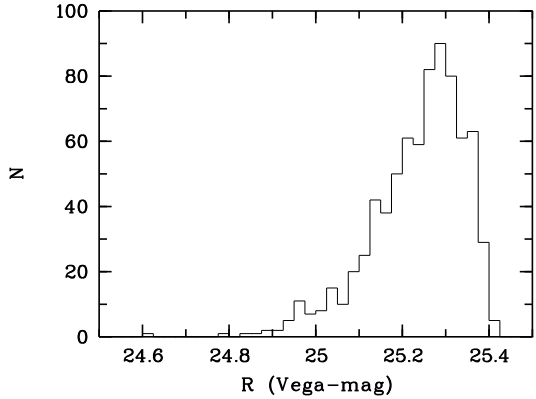,angle=270,clip=t,width=7cm}}}
\caption[ ]{R-band histogram of objects with errors of $\sim 10\%$ from combined
photometry of both observing runs. From this distribution we read the 10$\sigma$ 
limiting magnitude as the median at $R=25\fm25$. \label{errhist}}
\end{figure}

Fig.\,\ref{errors} shows the photometric errors versus magnitudes of
all objects we measured at less than $0\fm1$ error. The photon noise limit 
can be seen as a sharp parabolic edge to the right of the object clouds. 
We use magnitude histograms of objects with errors of $\sim 10\%$ to assess 
a representative 10-$\sigma$ 
magnitude limit for point-source photometry. In Fig.\,\ref{errhist} we 
can see that all $R$-band images combined reach $R_{\mathrm{lim, 10}\sigma} 
= 25.25$ which is $R_{\mathrm{lim, 5}\sigma} = 26.00$.

\begin{figure}
\centerline{\hbox{
\psfig{figure=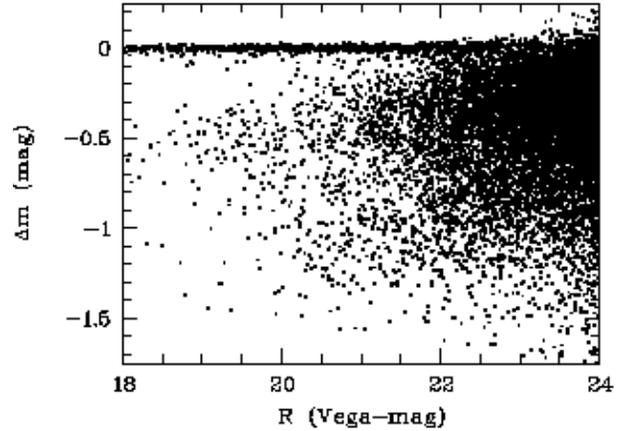,angle=270,clip=t,width=8cm}}}
\caption[ ]{Distribution of aperture correction magnitudes versus
R magnitude. Stars are at zero level while extended objects reach
down to negative values. \label{Apercorr}}
\end{figure}

\begin{figure}
\centerline{\hbox{
\psfig{figure=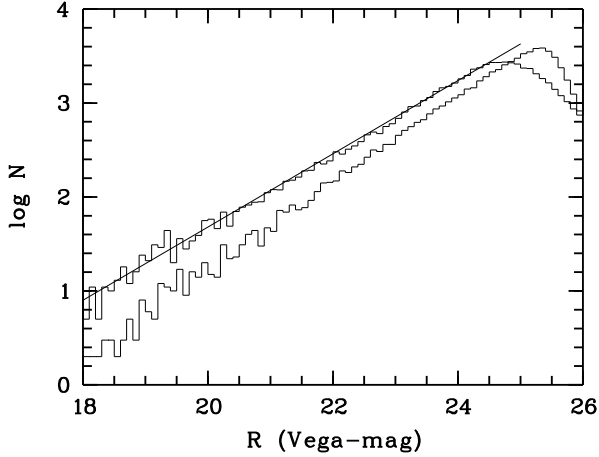,angle=270,clip=t,width=8cm}}}
\caption[ ]{Number counts histogram of all objects excluding stellar 
objects with $R<23$. The lower staircase line is based on the aperture magnitude
which is accurately calibrated for point sources. The upper staircase line is 
based on total magnitudes corrected for aperture effects, while the straight black
line with a slope of 0.39 has been fitted to the latter staircase distribution. Note, 
that this plot contains $\sim 62000$ objects and shows their counts in bins of $0\fm1$. \label{galNC}}
\end{figure}

However, the fluxes of extended objects are approximated by an aperture correction. 
The median correction among extended objects amounts to $\sim -0\fm5$ (see Fig.\,\ref{Apercorr}).
Fig.\,\ref{galNC} shows a number counts histogram for the aperture magnitude
as well as for the corrected total magnitude excluding point sources at $R<23$. 
The counts in aperture magnitude suggest that our object list is at least complete
to $R>25$ in terms of point-source photometry. The counts of total magnitude suggest
completeness among galaxies provided to at least $R\approx 24.5$. With a slope of 
$\sim 0.39$ the counts are consistent with galaxy counts from the literature.
However, a detailed discussion of the counts is beyond the scope of this paper.

\begin{figure}
\centerline{\hbox{
\psfig{figure=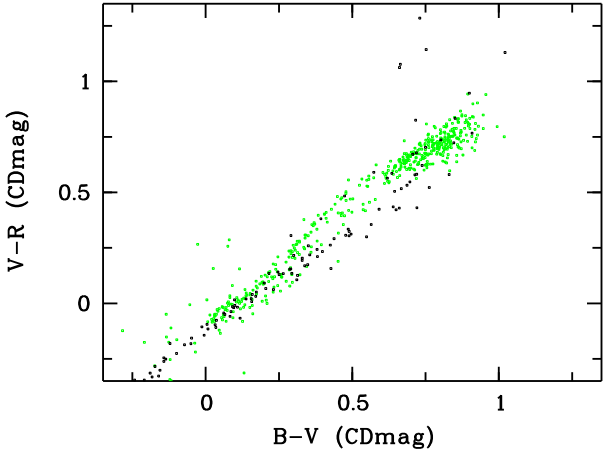,angle=270,clip=t,width=8cm}}}
\caption[ ]{$B-V$ and $V-R$ colors of observed stars compared to template
colors. Shown are bright point sources from the object catalogue ($R=17\ldots 21$,
grey dots) and colors for the Pickles (1998) spectral library (black dots). 
Most stars in our field belong to the halo population, while most of the Pickles 
library are nearby stars from the disk population. The populations form two arms 
that are clearly separated for G and K stars ($B-V \approx 0.3 \ldots 0.6$). 
We note a number of blue objects off the main sequence, which are probably
quasars passing this purely morphological selection. The units of the
colors are photon count color indices (CDmag) as in Wolf et
al. (2001), see also text. \label{ms}}
\end{figure}

We checked our flux calibration by comparing measured colors of stellar
objects with those predicted by synthetic photometry. We convolved the 
Pickles (1998) library of stellar spectra with the total efficiency curves 
of our filters and plotted their $B-V$ and $V-R$ colors as black dots in 
Fig.\,\ref{ms}. Our own point sources are overplotted as grey dots and 
agree with the expected colors without any further correction. Shown are
photon count color indices in units of CDmag defined by Wolf et al. (2001). 
As a physical magnitude definition besides ABmag and STmag the CDmag is
\begin{eqnarray}
& CD\mbox{mag} = & -2.5\log {F_{\mathrm{phot}}} + 20\fm01 \\ & & \mbox{with} ~
F_{\mathrm{phot}} ~ \mbox{in $\gamma$\,m$^{-2}$\,s$^{-1}$\,nm$^{-1}$} \nonumber
~,
\end{eqnarray}
and matches the common magnitude zeropoint of (astronomical) Vega-mag,
ABmag and STmag at $\lambda_0 = 548$\,nm.

We note that most stars observed in our field belong to the halo
population, while most stars in the Pickles library are nearby stars
from the disk population. The populations form two arms that are 
separated for G/K stars ($B-V_{CD} \approx 0.3 \ldots 0.6$) and have 
quite different relative population densities in the data and the library.
We also note a few blue objects off the main sequence, which
are most likely quasars passing the purely morphological selection used here.

\section{A first set of variable objects}\label{vars}

\begin{figure}
\centerline{\hbox{
\psfig{figure=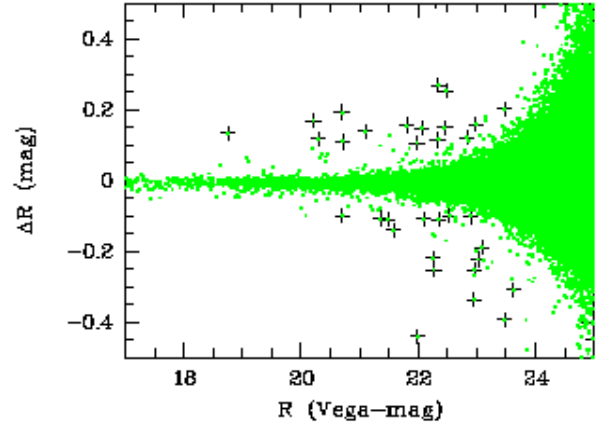,angle=270,clip=t,width=8cm}}}
\caption[ ]{$R$-band magnitude difference between the two epochs October
1999 and February 2000 versus combined $R$-magnitude (grey dots = 
all objects). Objects selected as variable are shown as black
crosses, but strongly variable sources lie outside the plot. Twenty
objects have been omitted from the sample after
visual inspection since they were affected by scattered light or
diffraction spikes of $9^{th}$ magnitude stars. \label{varmagdiff}}
\end{figure}

\begin{table*}
\caption{A first sample of variable objects drawn from $R$-band photometry of two epochs. While most objects are likely to be quasars at $z<3$, some objects might be Seyfert galaxies or even stars. We marked the objects where we assume a Supernova has caused the variability. The object numbers refer to the published catalog and the object names encode the object positions. \label{vartab} }
\begin{tabular}{rllllllll}
cat. nr. & object name & $\alpha_{\mathrm{J2000}}$ & $\delta_{\mathrm{J2000}}$ 
   & $B_{\mathrm{Oct99}}$ & $V_{\mathrm{Oct99}}$ & $R_{\mathrm{Oct99}}$ & $R_{\mathrm{Feb00}}$ & comment \\ 
\noalign{\smallskip} \hline \noalign{\smallskip} 
     4050 & COMBO--0332203$-$280215  &  $03^{\mathrm{h}} 32^{\mathrm{m}} 20\fs 315$  &  $-28\degr 02\arcmin 14\farcs66$  &  $21.146$  &  $21.013$  &  $20.698$  &  $20.605$ \\
     4427 & COMBO--0331278$-$280156  &  $03^{\mathrm{h}} 31^{\mathrm{m}} 27\fs 766$  &  $-28\degr 01\arcmin 55\farcs96$  &  $24.537$  &  $24.213$  &  $23.084$  &  $22.903$ \\
     4809 & COMBO--0331362$-$280150  &  $03^{\mathrm{h}} 31^{\mathrm{m}} 36\fs 247$  &  $-28\degr 01\arcmin 49\farcs59$  &  $22.695$  &  $22.525$  &  $22.258$  &  $22.046$ \\
     5826 & COMBO--0332470$-$280119  &  $03^{\mathrm{h}} 32^{\mathrm{m}} 46\fs 953$  &  $-28\degr 01\arcmin 19\farcs38$  &  $24.805$  &  $23.736$  &  $22.906$  &  $22.812$ \\
     7139 & COMBO--0332354$-$280041  &  $03^{\mathrm{h}} 32^{\mathrm{m}} 35\fs 368$  &  $-28\degr 00\arcmin 41\farcs19$  &  $24.338$  &  $24.225$  &  $23.629$  &  $23.328$ \\
     7902 & COMBO--0332302$-$280020  &  $03^{\mathrm{h}} 32^{\mathrm{m}} 30\fs 200$  &  $-28\degr 00\arcmin 19\farcs95$  &  $22.308$  &  $21.901$  &  $21.809$  &  $21.972$ \\
    13244 & COMBO--0333148$-$275749  &  $03^{\mathrm{h}} 33^{\mathrm{m}} 14\fs 849$  &  $-27\degr 57\arcmin 48\farcs96$  &  $23.992$  &  $24.026$  &  $23.473$  &  $23.087$ \\
    15278 & COMBO--0331208$-$275649  &  $03^{\mathrm{h}} 31^{\mathrm{m}} 20\fs 762$  &  $-27\degr 56\arcmin 48\farcs97$  &  $20.983$  &  $20.829$  &  $20.719$  &  $20.835$ \\
    15396 & COMBO--0332161$-$275644  &  $03^{\mathrm{h}} 32^{\mathrm{m}} 16\fs 133$  &  $-27\degr 56\arcmin 44\farcs12$  &  $23.064$  &  $22.916$  &  $22.533$  &  $22.440$ \\
    16155 & COMBO--0332042$-$275626  &  $03^{\mathrm{h}} 32^{\mathrm{m}} 04\fs 157$  &  $-27\degr 56\arcmin 26\farcs39$  &  $24.130$  &  $23.860$  &  $23.029$  &  $22.814$ \\
    16404 & COMBO--0332226$-$275622  &  $03^{\mathrm{h}} 32^{\mathrm{m}} 22\fs 604$  &  $-27\degr 56\arcmin 22\farcs48$  &  $24.001$  &  $23.360$  &  $22.339$  &  $22.616$ & SN candidate \\
    20787 & COMBO--0333053$-$275409  &  $03^{\mathrm{h}} 33^{\mathrm{m}} 05\fs 312$  &  $-27\degr 54\arcmin 09\farcs11$  &  $22.525$  &  $21.951$  &  $21.576$  &  $21.442$ \\
    27080 & COMBO--0333042$-$275103  &  $03^{\mathrm{h}} 33^{\mathrm{m}} 04\fs 206$  &  $-27\degr 51\arcmin 02\farcs77$  &  $24.619$  &  $23.535$  &  $22.332$  &  $22.455$ \\
    28275 & COMBO--0331525$-$275027  &  $03^{\mathrm{h}} 31^{\mathrm{m}} 52\fs 508$  &  $-27\degr 50\arcmin 27\farcs48$  &  $24.062$  &  $23.586$  &  $22.456$  &  $22.613$ \\
    29793 & COMBO--0333104$-$274945  &  $03^{\mathrm{h}} 33^{\mathrm{m}} 10\fs 389$  &  $-27\degr 49\arcmin 44\farcs75$  &  $24.112$  &  $23.775$  &  $23.202$  &  $22.051$ \\
    30792 & COMBO--0332432$-$274914  &  $03^{\mathrm{h}} 32^{\mathrm{m}} 43\fs 239$  &  $-27\degr 49\arcmin 14\farcs11$  &  $22.442$  &  $22.474$  &  $22.288$  &  $22.801$ \\
    32254 & COMBO--0333263$-$274831  &  $03^{\mathrm{h}} 33^{\mathrm{m}} 26\fs 301$  &  $-27\degr 48\arcmin 31\farcs14$  &  $23.329$  &  $23.352$  &  $22.957$  &  $22.712$ \\
    34357 & COMBO--0332087$-$274734  &  $03^{\mathrm{h}} 32^{\mathrm{m}} 08\fs 669$  &  $-27\degr 47\arcmin 34\farcs20$  &  $19.422$  &  $19.141$  &  $18.765$  &  $18.907$ \\
    35677 & COMBO--0332142$-$274647  &  $03^{\mathrm{h}} 32^{\mathrm{m}} 14\fs 246$  &  $-27\degr 46\arcmin 47\farcs44$  &  $24.293$  &  $24.051$  &  $23.477$  &  $23.685$ \\
    36683 & COMBO--0333343$-$274621  &  $03^{\mathrm{h}} 33^{\mathrm{m}} 34\fs 343$  &  $-27\degr 46\arcmin 20\farcs94$  &  $25.560$  &  $23.748$  &  $23.329$  &  $25.408$ & SN candidate \\
    37487 & COMBO--0332391$-$274602  &  $03^{\mathrm{h}} 32^{\mathrm{m}} 39\fs 084$  &  $-27\degr 46\arcmin 01\farcs82$  &  $21.101$  &  $20.944$  &  $20.696$  &  $20.897$ \\
    38551 & COMBO--0332300$-$274530  &  $03^{\mathrm{h}} 32^{\mathrm{m}} 29\fs 982$  &  $-27\degr 45\arcmin 29\farcs84$  &  $21.692$  &  $21.480$  &  $21.102$  &  $21.249$ \\
    38905 & COMBO--0333036$-$274519  &  $03^{\mathrm{h}} 33^{\mathrm{m}} 03\fs 616$  &  $-27\degr 45\arcmin 18\farcs71$  &  $23.744$  &  $23.362$  &  $22.833$  &  $22.962$ \\
    39432 & COMBO--0332302$-$274505  &  $03^{\mathrm{h}} 32^{\mathrm{m}} 30\fs 218$  &  $-27\degr 45\arcmin 04\farcs54$  &  $22.545$  &  $22.344$  &  $21.977$  &  $22.089$ \\
    41159 & COMBO--0332109$-$274415  &  $03^{\mathrm{h}} 32^{\mathrm{m}} 10\fs 924$  &  $-27\degr 44\arcmin 14\farcs73$  &  $23.073$  &  $22.931$  &  $22.350$  &  $22.247$ \\
    41247 & COMBO--0331534$-$274412  &  $03^{\mathrm{h}} 31^{\mathrm{m}} 53\fs 360$  &  $-27\degr 44\arcmin 12\farcs11$  &  $24.048$  &  $23.752$  &  $22.952$  &  $22.622$ \\
    41776 & COMBO--0333358$-$274400  &  $03^{\mathrm{h}} 33^{\mathrm{m}} 35\fs 841$  &  $-27\degr 44\arcmin 00\farcs07$  &  $23.408$  &  $23.287$  &  $22.981$  &  $23.145$ \\
    42601 & COMBO--0332591$-$274340  &  $03^{\mathrm{h}} 32^{\mathrm{m}} 59\fs 067$  &  $-27\degr 43\arcmin 39\farcs53$  &  $22.463$  &  $22.176$  &  $21.482$  &  $21.380$ \\
    43151 & COMBO--0332004$-$274319  &  $03^{\mathrm{h}} 32^{\mathrm{m}} 00\fs 365$  &  $-27\degr 43\arcmin 19\farcs45$  &  $22.726$  &  $22.436$  &  $22.062$  &  $22.216$ \\
    46562 & COMBO--0332479$-$274148  &  $03^{\mathrm{h}} 32^{\mathrm{m}} 47\fs 915$  &  $-27\degr 41\arcmin 47\farcs88$  &  $23.254$  &  $22.887$  &  $22.096$  &  $21.996$ \\
    47501 & COMBO--0331187$-$274121  &  $03^{\mathrm{h}} 31^{\mathrm{m}} 18\fs 698$  &  $-27\degr 41\arcmin 21\farcs30$  &  $22.768$  &  $22.291$  &  $21.989$  &  $21.556$ \\
    48870 & COMBO--0333289$-$274044  &  $03^{\mathrm{h}} 33^{\mathrm{m}} 28\fs 941$  &  $-27\degr 40\arcmin 43\farcs68$  &  $22.550$  &  $22.537$  &  $22.477$  &  $22.738$ \\
    52103 & COMBO--0331256$-$273908  &  $03^{\mathrm{h}} 31^{\mathrm{m}} 25\fs 611$  &  $-27\degr 39\arcmin 08\farcs48$  &  $25.298$  &  $24.421$  &  $23.274$  &  $24.475$ & SN candidate \\
    52280 & COMBO--0333211$-$273912  &  $03^{\mathrm{h}} 33^{\mathrm{m}} 21\fs 079$  &  $-27\degr 39\arcmin 11\farcs92$  &  $20.691$  &  $20.529$  &  $20.212$  &  $20.388$ \\
    57527 & COMBO--0333184$-$273641  &  $03^{\mathrm{h}} 33^{\mathrm{m}} 18\fs 358$  &  $-27\degr 36\arcmin 41\farcs34$  &  $22.785$  &  $22.331$  &  $21.360$  &  $21.260$ \\
    58758 & COMBO--0333037$-$273611  &  $03^{\mathrm{h}} 33^{\mathrm{m}} 03\fs 722$  &  $-27\degr 36\arcmin 10\farcs90$  &  $20.924$  &  $20.755$  &  $20.298$  &  $20.423$ \\
    59821 & COMBO--0332443$-$273403  &  $03^{\mathrm{h}} 32^{\mathrm{m}} 44\fs 276$  &  $-27\degr 34\arcmin 03\farcs31$  &  $22.741$  &  $22.564$  &  $22.265$  &  $22.019$ \\
\noalign{\smallskip} \hline
\end{tabular}
\end{table*}

For variability studies we calculated a magnitude change between the
two epochs October 1999 and February 2000 and a related error as:
\begin{equation}
\Delta R = R_{\mathrm{Feb00}} - R_{\mathrm{Oct99}} ~,
\end{equation}

and
\begin{equation}
\sigma_{\mathrm{\Delta R}} = \sqrt{\sigma_{R_{\mathrm{Oct99}}}^2 + \sigma_{R_{\mathrm{Feb00}}}^2}
~.
\end{equation}

Fig.\,\ref{varmagdiff} shows the magnitude difference of all
objects with $R<25$ versus the combined $R$-magnitude. The distribution
underlines that the photometric errors are realistic and the
photometry is essentially consistent among the two epochs.

We selected a first sample of variable objects using the following criteria:
\begin{enumerate}
\item faint cutoff: at least one detection must have $R<24$
\item absolute variability threshold: $\Delta R > 0.1$
\item variability significance threshold: $\Delta R /\sigma_{\Delta R}
> 5$
\end{enumerate}

This way 57 entries are selected in the catalogue which were made
subject to visual inspection on the sum frames of the two epochs. We
eliminated 20 faint objects from the sample which were affected by scattered
light or diffraction spikes of $9^{th}$ magnitude stars. We have
confidence in the variability of the remaining 37 objects and list
their positions and original photometry from the two epochs in
Table\,\ref{vartab}. The number of variable objects increases strongly
with fainter magnitude but faintwards of $R \sim 22.5$ the
significance criterion becomes stronger than the absolute threshold,
thereby reducing the numbers again. The published catalogue of course 
provides the possibility to identify objects with a smaller amplitude
of variability.

\begin{figure}
\centerline{\hbox{
\psfig{figure=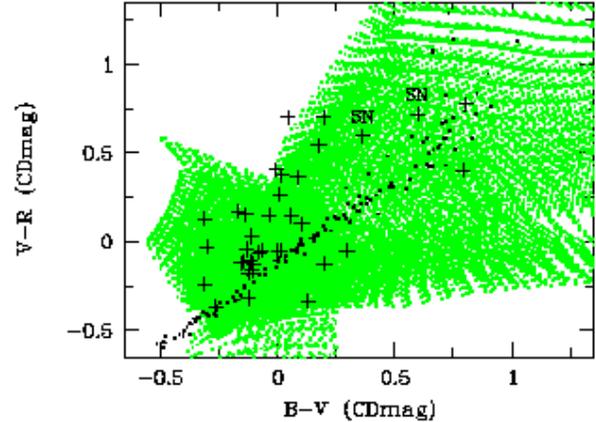,angle=270,clip=t,width=8cm}}}
\caption[ ]{Color-color diagram of variable objects (black crosses)
compared with the expected stellar sequence (black dots) and a quasar
color library (grey dots). Assuming the absence of short-term
variability, most objects would be AGNs. Two Supernova (SN) 
candidates are within the limits of this diagram. \label{varcolors}}
\end{figure}

We show the location of the variables in a $(B-V)$ vs.\ $(V-R)$ color diagram in
Fig.\,\ref{varcolors}. The red magnitude is here only taken from
October 1999 images to keep the color indices free from being skewed
by long-term variability. The October frames have been taken during 9 days,
so short-term variability could still have changed the color indices,
especially for RR Lyrae stars given their typical periods of $\sim 1$
day. Also shown are expected colors for
stars from the Pickles atlas (compare with Fig.\,\ref{ms}) and for a
quasar color library \cite{WMR01} derived from quasar spectra modelled
for $z=0\ldots6$ as a combination of power-law spectra, an
emission-line contour \cite{Fra91} and Lyman-forest.

If we assume an absence of short-term variability, we would classify
most objects as quasars or Seyfert galaxies, while a few could be stars. 
Of course, the stellar sample could contain some RR Lyrae stars whose 
variability time scale of $\sim 1$ day could offset their colors. 
These color offsets could move the stars in any directions (+/-) along 
any of the color axes with equal probability, and a scattered sample 
should appear with a comparable fraction of stars below and above the 
original color sequence.

In Fig.\,\ref{varcolors} we see that only three objects show up below
the stellar sequence, while $\sim 30$ are located above. Therefore we
conclude that most of the variable objects presented are indeed
quasars while some objects are Seyfert-I galaxies where bright host 
galaxies dominate the colors. The COMBO-17 group has meanwhile reduced 
all data on the Chandra field and prepared lists of stars, galaxies 
and quasars at $R \la 24$, which are analysed in forthcoming papers.

Among the variable objects are also three candidates for Supernovae
(see Table\,\ref{vartab}), which had their bright phase in October
1999 and are undetectable to visual inspection in the February coadded
frame. The first two showed up as a significant off-center brightening 
of a tiny faint galaxy (objects 16404 and 52103 in Table\,\ref{vartab}).

The third candidate has a rather red $B-V$ color (object 36683 in
Tab.\,\ref{vartab}). It is a point source located only $2\farcs0$
North and $0\farcs9$ East of a small galaxy with $R=22.93\pm0.02$ 
(no calibration errors included). There is still non-zero flux measured 
at the location of the transient object in February 2000 formally yielding 
$R=25.41\pm0.19$, but it is not clear, whether the light originates
from the outskirts of the neighboring galaxy or still from the
Supernova.

In October it was measured on Oct 10 with $V=23.75\pm0.04$, on Oct 13
with $B=25.56\pm0.17$ and on Oct 19 and 20 consistently with
$R=23.33\pm0.02$. Again, we can not decide on the B band sum frame
with $1\farcs2$ PSF whether the B flux is contributed from the galaxy
or from the candidate. If it was a Supernova, the B measurement 
bracketed by V and R imaging could not be explained by variability. 
Therefore, the object must be unusually red with $B-V \ga 1.8$ and
it is not entirely clear whether the Supernova hypothesis can
account for this color.

We note, that the Supernova 1999gu reported by Cappellaro et al. (2000) 
and first observed with the WFI on Dec 29, 1999 at $\alpha_{\mathrm{J2000}} 
= 3^{\mathrm{h}}33^{\mathrm{m}}00\fs1$ and $\delta_{\mathrm{J2000}} =
$-$27\degr51\arcmin40\arcsec$ in a galaxy at $z=0.147$ is clearly
visible in the images of February 2000 with $R=21.5$, but since our
object list was defined on the deep sum frame of October 1999, it is
not contained in the catalogue presented here.

\section{Summary and Outlook}

We have constructed a catalogue with positions, morphology and deep $BVR$
photometry ($B$,$V_{5\sigma}\approx 25.5$ $R_{5\sigma} = 26$) of 
63501 objects on an area of $31\farcm5 \times 30\arcmin$
containing the Chandra Deep Field South. This catalogue is available to
the scientific public at Centre de Données astronomiques de Strasbourg
(CDS, \texttt{http://cdsweb.u-strasbg.fr/}). 

We have presented a first list of faint variable objects, which are 
supposedly mostly quasars including some Seyfert galaxies or Supernovae
in late stages. Three transient sources are strongly suggestive of
supernovae, but one of them had an unusually red color of $B-V \ga 1.8$.

When fully reduced, the dataset collected by the COMBO-17 survey will
provide a bonanza of pseudo-spectroscopic information. 
We expect to classify some 50\,000 objects over an area of 1~deg$^2$
down to $R \la 24$ (completeness limit). Besides the classification 
infomation (star, galaxy or quasar) we will get spectral subclasses and 
high-quality redshift estimates for extragalactic sources. 
The full catalogue will then allow to finally classify also the variable
sources and tell the redshifts of the Supernova host galaxies. It will
not only provide an optical classification of the X-ray sources but
also of neighboring objects in their environment.

\begin{acknowledgements}
This work was supported by the DFG (Sonderforschungsbereich 439). 
\end{acknowledgements}

\end{document}